\documentclass[letterpaper,twocolumn,10pt]{article}
\usepackage{usenix,epsfig,endnotes,color,verbatim,listings,hyperref,bigstrut,multirow}
\begin{document}
\newcommand{\murl}[1] {{\sf \url{#1}}}
\lstdefinelanguage{patch}{morekeywords={p,-,+,=},
  keywordstyle=\small\bf\sffamily,
  basicstyle=\small\sffamily,
  morecomment=[n]{/*}{*/},
  morecomment=[l]{//},
}

\newcommand{\aspa} {{\sf aspa} }
\newcommand{\xdelta} {{\sf xdelta} }
\newcommand{\diff} {{\sf diff} }

\newcommand{\psection}[1] {\vspace{0.2cm}\noindent {\bf #1}. }
\date{}

\title{\Large \bf Fine-grained patches for Java software upgrades}

\author{
{\rm Eduardo R. B. Marques}\\
Faculdade de Ci\^{e}ncias da Universidade de Lisboa \\
{\sf edrdo@di.fc.ul.pt}
} 

\maketitle


\subsection*{Abstract}

We present a novel methodology for deriving fine-grained patches of Java software. We consider an abstract-syntax tree
(AST) representation of Java classes compiled to  the 
Java Virtual Machine (JVM) format, and a difference
analysis over the AST representation to derive patches.
The AST representation defines an appropriate abstraction level 
for analyzing differences, yielding compact patches
that correlate  modularly to actual source code changes. 
The approach contrasts to other common, 
coarse-grained approaches, like plain binary differences, 
which may easily lead to disproportionately large patches.
We present the main traits of the methodology, a prototype tool
called \aspa that implements it, and a case-study analysis 
on the use of \aspa to derive patches for the Java~2~SE API.
The case-study results illustrate that \aspa patches  
have a significantly smaller size than patches 
derived by binary differencing tools.

\section{Introduction}

Echoing Lehman's law of continuous change~\cite{lehman}, 
modern software is evolving constantly and  software upgrades are routinely deployed. 
Consider for instance ``smartphone apps'', where upgrades are very common 
and directly affect end user experience
in many ways, like data transfer and associated cost, installation time,
or  user intervention.
Thus, software upgrades must be increasingly reliable, efficient, and automated,  both for the end user and the software vendor or provider. 

Upgrades are defined by software \emph{patches}, reflecting the transition between software versions. A patch~$p$ between  old and new versions~$O$ and~$N$ of a software artifact in compiled form is such that $N = p(O)$.
That is, the patch~$p$ must encode the transformation from~$O$ to~$N$, that occurs as part of the upgrade from~$O$ to~$N$ in a target platform.
It is many times the case that~$p$ is not an incremental transformation of~$O$,  but instead 
amounts to the entire $N$, i.e., $p = N$, 
corresponding to the full installation of the new version,
e.g., as in Android smartphone apps.
More refined, incremental approaches define~$p$ 
as the set of changed files from~$O$ to~$N$, 
or attend to the binary differences
between~$O$ and~$N$ or between component files within~$O$ 
and~$N$~\cite{courgette,bsdiff,percival,win-bdc}.

All of the above approaches are common. The problem is that they
are too coarse-grained and operate at an inappropriate abstraction level.
In the general case, a resulting patch~$p$ may not  appropriately reflect the may have a disproportionate size
to the
``actual'' changes between~$O$ and~$N$, 
i.e., the relevant syntactical differences between $O_{\sf S}$ and $N_{\sf S}$.

We propose a better solution to 
this problem, in the context of Java software~\cite{jls} compiled 
to the Java Virtual Machine~\cite{jvm} bytecode format.
The approach is to account for the \emph{fine-grained
changes} between two versions~$O$ and~$N$ of a JVM class file, 
expressed in an abstract syntax tree (AST) representation. 
The JVM class file format is closely related to the core traits of Java
in source form. Hence JVM-level patches 
may potentially  correlate more evenly with source-code level  changes,
whilst avoiding the obvious inconveniences of using source-code based patches
for upgrades (e.g., a Java compiler on the target platform, IP issues). 
On the other hand, since the JVM format is for all purposes still
a binary one, an AST representation may factor out features
that are syntactically irrelevant or do not correlate
to source code changes, e.g., the definition order of 
methods in a JVM class file or constant pool indexes spread
throughout it~\cite{jvm}.

We have developed a prototype tool called \aspa that implements this
methodology, written in Java and available  from~\cite{web}.
In the remainder of the paper, we begin by the describing the main traits of the methodology and the \aspa tool (Section~\ref{sec:deriving}). We then present results of using \aspa over the core Java~2~SE~(J2SE)~API, showing 
that \aspa patches can be much smaller than binary difference patches
(Section~\ref{sec:results}).
We end the paper with a discussion of related work, highlights for future work, and possible use of the presented methodology 
in other contexts~(~Section~\ref{sec:conclusion}).

\vspace{-0.3cm}
\section{The \aspa tool}\label{sec:deriving}
\vspace{-0.2cm}

\newcommand{\Cold} {C_{{\rm O}}}
\newcommand{\Cnew} {C_{{\rm N}}}
\newcommand{\Aold} {A_{{\rm O}}}
\newcommand{\Anew} {A_{{\rm N}}}
\newcommand{\fAST} {{\sf ast}}
\newcommand{\fdiff} {{\sf diff}}
\newcommand{\fjvm} {{\sf jvm}}

\begin{figure}[htbp!]
\centering
\includegraphics[scale=1.25]{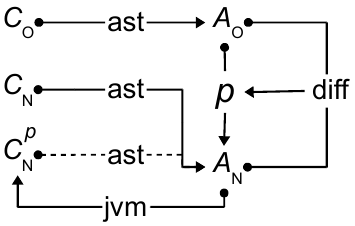}
\caption{\label{overview}The \aspa approach}
\end{figure}

\psection{Overview} Our approach is illustrated in~Fig.~\ref{overview}.  Given old and new versions of a Java class in the binary JVM format, $\Cold$ and~$\Cnew$, the 
\aspa tool parses both  to derive corresponding symbols $\Aold~=~\fAST(\Cold)$ and~$\Anew~=~\fAST(\Cnew)$ with an AST-like representation. A patch for the upgrade from~$\Cold$ to~$\Cnew$ 
is derived by computing AST-level differences between~$\Aold$ and~$\Anew$,
~$p=~\fdiff~(\Aold,~\Anew)$. The patch~$p$ can then be applied to~$\Aold$ to obtain~$\Anew$, i.e.,~$\Anew=p(\Aold)$, after which~$\Anew$ can be converted back to the JVM format,~$\Cnew^p~=~\fjvm(\Anew)$.

\newcommand{\keyE}[1] {{\bf #1}}
\newcommand{\entryA}[2] { ${\sf #1}$ & ${\sf ::=}$ & ${\sf #2}$  \\ \bigstrut}
\newcommand{\entryB}[1] { & & ${\sf #1}$  \\ \bigstrut }

\begin{figure}[h!]
\begin{tabular}{@{}r@{ }c@{ }l@{}}
\entryA{Class}{\keyE{class}:ObjectType}
\entryB{superclass:ObjectType}
\entryB{interfaces:set(ObjectType)}
\entryB{fields:set(Field)}
\entryB{methods:set(Method)}
\entryB{pool:set(Constant)}
\entryB{attributes:set(Attribute)}
\entryB{version:i4} 
\entryB{flags:i2} 
\entryA{Constant}{i4\ |\ i8\ |\ f4\ |\ f8\ |\ utf8\ |\ ... }
\entryA{Field}{\keyE{name}:utf8}
  \entryB{type:Type} 
  \entryB{flags:i2}
  \entryB{attributes:set(Attribute)}
\entryA{Method}{\keyE{signature}:Signature}
 \entryB{flags:i2}
 \entryB{attributes:set(Attribute)}
 \entryA{Signature}{name:utf8}
    \entryB{return:Type} 
    \entryB{args:seq(Type)}
\entryA{Attribute}{\keyE{name}:utf8}
 \entryB{content:AttrType}
\entryA{AttrType}{Code \ |\  Exceptions\ |\ ... }
\entryA{Code}{instructions:seq(Instruction)}
  \entryB{max\_stack:u2}
  \entryB{locals:u2}
  \entryB{attributes:set(Attribute)}
\entryA{Instruction}{aload\_0\ |\ iconst\_m0  \ |\ ...} 
\entryA{Type}{Type[]\ |\ ObjectType\ |\ {\tt int}\ |\ ...}
${\tt ObjectType}$ & ${\tt ::=}$ & $\mbox{class or interface identifier}$ \\
${\tt i<n>}$ &  ${\tt ::=}$ & ${\tt n}\mbox{-byte integer}$  \\ 
${\tt u<n>}$ &  ${\tt ::=}$ & ${\tt n}\mbox{-byte unsigned integer}$  \\ 
${\tt f<n>}$ &  ${\tt ::=}$ & ${\tt n}\mbox{-byte floating point constant}$  \\ 
\entryA{utf8}{ \mbox{UTF-8 string} }
...
\end{tabular}
\vspace{-0.3cm}
\caption{\label{grammar}AST representation of Java classes}
\vspace{-0.3cm}
\end{figure}

Note that in Fig.~\ref{overview} we can have that 
$\Cnew^{p} \neq \Cnew$ in terms of the binary JVM format, 
but in any case $\fAST(\Cnew)~=~\fAST(\Cnew^{p})~=~\Anew$ at the AST level.
This is a by-product of the core trait of {\sf aspa}:
$\Aold$, $\Anew$ and~$p$ factor out a number of serialization aspects in the JVM binary format  that are syntactically irrelevant, but lead to disproportionate binary-level differences between $\Cold$ and~$\Cnew$, e.g., the indexes of constants in the JVM pool or method definition order~\cite{jvm}. Binary differences are
by definition sensitive to these aspects, but \aspa factors them 
by out by appropriate mechanisms in AST data representation and difference analysis, described next.


\newcommand{\GS}[1] {${\sf {#1}}$}
\psection{AST representation of JVM class files}
In Fig.~\ref{grammar} we depict a fragment
of the abstract syntax grammar embedded into \aspa to represent a
Java class, using semi-formal BNF notation.
The grammar abstracts the core symbolic information found in a JVM class file~\cite{jvm} and, in close relation,
also the fundamental traits of Java classes in source code form~\cite{jls}.

Given a JVM class file $C$, \aspa  derives $\fAST(C)$,
an instance (production) of the \GS{Class} root symbol
in the grammar of Fig~\ref{grammar}.
A \GS{Class} instance is a tuple with the following labelled attributes: the class 
type (\GS{class}); its superclass ({\GS{superclass});
the sets of implemented interfaces, fields, methods, constants, and attributes (as shown);
the JVM format version (\GS{version}); and a flag mask (\GS{flags},
representing access modifiers like ${\sf public}$ and other  properties~\cite{jvm}).

Other tuple symbols in the grammar of Fig.~\ref{grammar} have a similar definition to \GS{Class}, such 
as \GS{Field} or \GS{Method}. For tuple symbols~$S$ such as these, attributes~$k$
shown in bold identify that two instances of $S$ should only be compared
(analyzed for differences) if they have the same value for~$k$, and are called
symbol keys. For instance, 
two instances of the \GS{Method} symbol should only be compared 
if they have the same value for the \GS{signature} attribute, 
i.e., two methods
are comparable if they have the same signature 
(same name, return type, argument count and types).
Also for tuple symbols, we use notation~${\sf seq}(S)$ and~${\sf set}(S)$ for some attributes in correspondence to 
sequences and sets of instances of symbol~$S$, respectively,
for instance~\GS{instructions:seq(Instruction)} in~\GS{Code}
and~\GS{methods:set(Method)} in~\GS{Class}.
The grammar is completed by terminal symbol derivations,
as shown for Java bytecode instructions (\GS{Instruction})
and constant values (integers, UTF-8 strings, etc).

\begin{figure}[htbp!]
\vspace{-0.3cm}
\begin{lstlisting}[language=Java]
// Old version
package toy;
class Foo {
 private int x;
 public Foo(){ x = 0; }
 public int sqX(){ return x * x; } 
 public int getX(){ return x; }
}
// New version
package toy;
class Foo {
 // y added
 private int x, y; 
  // sqX moved, but unchanged
 public int sqX(){ return x * x; } 
 // constructor changed
 public Foo(){ x = 1; y = 0; }    
 // getX removed
 // setX added
 public void setX(int v){ x = v; } 
} 
\end{lstlisting}
\vspace{-0.3cm}
\caption{\label{toy-source}Toy example --- Java source code}
\vspace{-0.1cm}
\end{figure}

\begin{figure}[h]
\begin{lstlisting}[language=patch]
p Foo {
 p constants {
   + utf8 "y"
   - utf8 "getX"
   + utf8 "setX"
   ...
 }
 p fields { 
   + name=y type=int flags=... 
 }
 p methods {
  p Foo() {
   p attributes {
    ...
    p Code { 
     p instructions { 
      = aload_0       
      = invokespecial java/lang/Object()
      = aload_0       
      - iconst_0
      + iconst_1      
      = putfield x
      + aload_0       
      + iconst_0      
      + putfield y
      = return 
    }
   }
  }
  - int getX() 
  + void setX(int) { ... omitted ... }
 }
}
\end{lstlisting}
\vspace{-0.3cm}
\caption{\label{toy-patch}Toy example --- derived patch}
\vspace{-0.3cm}
\end{figure}

The conversion from JVM to AST form abstracts a number of serialization
features that are syntactically irrelevant and help generating compact
patches, as opposed to being sensitive to the particular
layout of a JVM class file. Essentially, \aspa
factors out two main aspects. First, \aspa resolves
constant pool index references~\cite{jvm} at the AST level.
Constant pool indexes are spread
throughout the entire contents of a JVM class file and induce low-level
binary changes, when the index of a particular constant changes in-between 
software versions.
Secondly, the definition order of several symbols like fields, methods, etc 
is also factored out, as determined by the ${\sf set}(S)$ definitions in Fig.~\ref{grammar}. Hence, for instance, \aspa will consider two class files 
to be equivalent if they only differ by  the use of different JVM pool indexes for constants, or the order of definition of methods or fields.

\newcommand{\Sold} {s_{{\sf O}}}
\newcommand{\Snew} {s_{{\sf N}}}

\psection{Example} 
We first illustrate the process of patch derivation intuitively using a toy example. 
Two versions of a class named {\sf Foo} are shown in Fig.~\ref{toy-source}
and a human-readable representation of the patch between the two versions is shown in Fig.~\ref{toy-patch}.  We omit  AST 
representations of old and new versions of {\sf Foo}, 
as they would repeat the source code traits,
and are in any case also implicit in the patch representation shown.
The changes from old to new version are 
annotated in Fig.~\ref{toy-source} as Java comments 
and in Fig.~\ref{toy-patch} by notation ${\sf =}$, ${\sf +}$, ${\sf -}$, and ${\sf p}$
in correspondence to 
unchanged, added, removed and patched (i.e., changed) sections of the AST, respectively.
The changes from old to new version of {\sf Foo} are then as follows: 

--- Field {\sf y} is added;  

--- Method {\sf getX} is removed,  method {\sf setX} is added, while 
unchanged method {\sf sqX} is unaccounted for by the patch, in spite of being defined in a different order;

--- The {\sf Foo} constructor method is patched: 
its JVM bytecode instructions  contain a different 
value to initialize field {\sf x} ({\sf 1} in place of {\sf 0}), 
plus new instructions to initialize field {\sf y};

--- Constants are added to or removed from the JVM constant pool
in relation to all other changes, as exemplified in Fig~\ref{toy-source} for
the UTF-8 constants  in the {\sf p constants} section
({\sf ``y''}, {\sf ``getX''} and {\sf ``setX''}).

\psection{Path derivation} 
As illustrated by the example in Fig.~\ref{toy-patch},
an \aspa patch is a type of tree-edit script~\cite{tree-edit} over
the  AST representation of a Java class.
Given two AST representations, 
$\Aold$ and $\Anew$, \aspa matches the structure of~$\Aold$ and~$\Anew$ and derives as a result the patch $p=\fdiff(\Aold,\Anew)$,
such that~$\Anew = p(\Aold)$.
Generally, to derive the patch~$p$ from~$\Sold$
to $\Snew$, where~$\Sold$ and~$\Snew$ are two instances of some symbol~$S$ in the AST grammar, \aspa proceeds in syntax-driven manner as follows:

--- If~$\Sold$ and~$\Snew$ are plain terminals 
(e.g., instances of \GS{Instruction}) and $\Sold\neq\Snew$  then
$p$ is expressed (fully) by $s_N$. If $\Sold=\Snew$, 
for this and all the cases below, we define~$p$ as the identity mapping (denoted by~{\sf =} in~Fig.~\ref{toy-patch});

--- If~$S$ is a tuple symbol $S=(a_1:S_1,\:\ldots\:,a_n:S_n)$, for instance  \GS{Class},
then the patch is also a tuple $p=(p_1,\:\ldots,\:p_n)$ where $p_i = \fdiff \left(\Sold(a_i), \Snew(a_i)\right)$ for~$i~=~1,~\:\ldots\:,~n$;

--- If~$s_{\sf O}$ and~$s_{\sf N}$ are instances of a set attribute ${\sf set}(S)$, such as
\GS{methods:set(Method)} in \GS{Class}, then~$p$
can be derived using a set difference analysis that takes into account
the key attribute~$k$ of~$S$ if defined (e.g.,~\GS{signature} in~\GS{Method}).
Changes, additions and removals  from~$s_{\sf O}$ to~$s_{\sf N}$ 
can be identified in this manner, and ``tree moves'' (i.e.,~definition order) can be factored out.
Note that changes account 
for possible elements in both $s_{\sf O}$ and~$s_{\sf N}$
with the same key value, but which differ in some manner otherwise,
 e.g., like the \GS{Foo} constructor patch 
within the \GS{p\ methods} section of Fig.~\ref{toy-patch};

--- Finally, if~$s_{\sf O}$ and~$s_{\sf N}$ are instances of
 of a sequence attribute ${\sf seq}(S)$, such as \GS{instructions:seq(Instruction)} in \GS{Code}, then~$p$ can be expressed as a shortest-edit script (SES) over 
the longest common subsequence (LCS) of~$s_{\sf O}$ and~$s_{\sf N}$~\cite{lcs-survey}.
The SES expresses a sequence of symbol additions, removals,
and changes from~$s_{\sf O}$ to~$s_{\sf N}$, and is derived
by \aspa using the LCS/SES algorithm described in~\cite{lcs}.

\psection{Patch application} Given an AST representation $\Aold$ of a Java and a patch $p=\fdiff(\Aold,\Anew)$ for some other AST representation~$\Anew$, $p$ can be applied to~$\Aold$ to 
yield~$\Anew$, i.e., $\Anew = p(\Aold)$. The procedure is symmetrical to that of patch derivation described
above, hence we omit details that would be repetitive. 
It should suffice to say \aspa changes~$\Aold$ in syntax-driven manner, accounting for the incremental changes defined by~$p$, resulting in~$\Anew$ at the end. 

\psection{Patch format}
The \aspa binary patch format uses special marks to denote
symbol changes, additions, removals, etc,
but otherwise simply adheres to the JVM format
to encode the AST representation in binary form.
For instance, if \aspa encounters
a method that has been added to a class, it serializes the method definitions (including JVM bytecode) using the JVM format to the patch file. When reading the  same patch file for application, that method will be converted (resolved) from the JVM format back to an AST form. This amounts to (reusing) the same mechanism 
to convert between entire  classes in JVM format
and corresponding AST representations.

\vspace{-0.3cm}
\section{Case-study}\label{sec:results}
\vspace{-0.2cm}

\hyphenation{aspa bsdiff jardiff jarpatch}
\newcommand{\jardiff} {{\sf jardiff.sh} }
\newcommand{\rtjar} {{\sf rt.jar} }
\newcommand{\bsdiff} {{\sf bsdiff} }

\psection{The J2SE API}
The J2SE API is bundled in the {\sf rt.jar} JAR archive of the Java Runtime Environment (JRE) distribution by Oracle. 
The archive contents include well-known J2SE API packages,
such as {\sf java.lang} or {\sf java.util}.
To conduct a case-study analysis, we downloaded all JRE Java 7 versions for Linux x64 and extracted the \rtjar archive from each of them.  The versions at stake comprise the initial JRE 7 release, plus all subsequent updates available from Oracle's J2SE homepage as of April 4, 2013: updates 1 to 7, 9 to 11, 13, 15, and 17 (updates 8, 12, 14, and 16 are not made available).

\psection{Patch derivation}
For each pair of successive JRE 7 releases,
we derived \aspa patches for \rtjar using the {\sf jardiff.sh} utility script included in the \aspa distribution~\cite{web}.
This script is able to produce a single patch file, reflecting the differences of all class files between two versions of a JAR file.
The derived patch can then be applied to the source version JAR
using the {\sf jarpatch.sh} script~\cite{web}.

For comparison, we also derived patches for \rtjar using the {\sf bsdiff}~\cite{bsdiff,percival} binary patching tool. The tool is a well-known one for this purpose. For instance, \bsdiff is embedded in Google's Courgette tool to produce Google Chrome patches~\cite{courgette}. 
We only refer to the comparison of \aspa with \bsdiff
because it is the binary patching tool we have tested that compares more favorably with {\aspa}.  The comparison of \aspa with other binary patching tools is reported in~\cite{web}. For instance,
the table shows that JRE update 1 changed (patched) 20 classes over the initial JRE release, added 3 new ones, and removed none.

We did two adjustments 
to make the comparison between \aspa and \bsdiff patches as balanced as possible. First, since \bsdiff employs built-in {\sf bzip2 -9}  compression for patches, we compressed \aspa patches in the same manner. Secondly, we ran \bsdiff 
not over the \rtjar archives directly, but over corresponding files containing the concatenation of all JVM class files in the \rtjar archive, ordered by package and class names. 
The latter aims to factor out too many dependencies of the JAR 
archive format itself in \bsdiff patches,
which \aspa is capable of dealing with comparable less impact.
We examine the sensitivity of \aspa and \bsdiff patches
to variations in the binary input format later in the text.

\begin{table}[ht!]
{\small
\begin{center}
\begin{tabular}{@{}|c|c|c|c|c|r|r|@{}}
\hline
V & {\sf p} & ${\sf +}$ & ${\sf -}$ & $\sum$ & 
 \aspa & \bsdiff \bigstrut \\ \hline \hline
u01 & 20 & 3 & 0 & 23 & 8.2 & 12.3 \bigstrut \\ \hline 
u02 & 91 & 6 & 0 & 97 & 36.7 & 67.3 \bigstrut \\ \hline 
u03 & 25 & 3 & 0 & 28 & 27.2 & 35.4 \bigstrut \\ \hline 
u04 & 431 & 61 & 3 & 495 & 156.1 & 292.4 \bigstrut \\ \hline 
u05 & 46 & 0 & 0 & 46 & 12.5 & 27.7 \bigstrut \\ \hline 
u06 & 365 & 17 & 205 & 587 & 108.9 & 161.1 \bigstrut \\ \hline 
u07 & 78 & 36 & 8 & 122 & 27.1 & 47.1 \bigstrut \\ \hline 
u09 & 55 & 14 & 0 & 69 & 21.9 & 32.8 \bigstrut \\ \hline 
u10 & 26 & 5 & 0 & 31 & 13.4 & 22.1 \bigstrut \\ \hline 
u11 & 3 & 0 & 0 & 3 & 2.3 & 3.5 \bigstrut \\ \hline 
u13 & 150 & 28 & 18 & 196 & 51.0 & 94.0 \bigstrut \\ \hline 
u15 & 24 & 2 & 1 & 27 & 13.2 & 17.6 \bigstrut \\ \hline 
u17 & 10 & 2 & 0 & 12 & 8.9 & 12.0 \bigstrut \\ \hline 
\end{tabular}
\end{center}
\vspace{0.1cm}
{\small
{\sf V}: JRE version for {\sf rt.jar}; {\sf p}: patched classes; 
${\sf +}$: added classes; ${\sf -}$: removed classes;
$\sum$: total number of patched/added/removed classes; 
{\sf aspa}: size of \aspa patch (KB);
{\sf bsdiff}: size of \bsdiff patch (KB)
}
}
\caption{\label{table}\rtjar class changes and patch sizes}
\vspace{-0.4cm}
\end{table}

\psection{Patch size comparison} 
We summarize in Table~\ref{table} the evolution of the \rtjar archive
between successive JRE 7 updates, and the sizes of corresponding \aspa and \bsdiff patches. The numbers shown for patched, added, and removed classes in each JRE 7 update were calculated by \aspa during patch derivation. 

The general conclusion to draw from Table~\ref{table} is that \aspa patches can be significantly smaller than \bsdiff patches. 
On average, the size of the derived \bsdiff patches was
$1.65$ times the size of \aspa patches,
from a minimum factor of $1.3$ (for {\sf u3} and {\sf u15}), to a maximum one of $2.2$ (for {\sf u05}).

We also measured the cross-correlation coefficient between the statistical distributions of total class changes 
(the~$\sum$ column in Table~\ref{table}) 
and the size of patches 
(the \aspa and \bsdiff columns). The coefficients are $0.94$ for~\aspa and $0.90$ for~{\sf \bsdiff}, indicating that \aspa patches seem to be in more modular correlation to actual variations between versions of the \rtjar archive. 

\psection{Binary patching sensitivity}
The size of binary patches is naturally sensitive to variations in the input format, and can be quite disproportionate to the actual changes
between software versions.
We illustrate this point with two examples:

1) If, similarly to {\sf aspa},
 the JAR files are provided directly as input to {\sf bsdiff}
(in place of ``flat'' JVM files, described previously),
 the patch size is increased significantly.
For instance,
the \bsdiff patch for the update between JRE versions {\sf u10} and {\sf u011}  (the smallest in size in Table~\ref{table}) 
grows from~3.5 KB to~79.12 KB. In contrast, \aspa
abstracts JAR file entry details (like CRC checksums or
timestamps) that are irrelevant.

2) The \bsdiff patches will even become larger, and by another order of magnitude,  if we use an \aspa generated JAR in place of the (AST-equivalent) JRE counterpart, given that 
\aspa follows its own JVM serialization
strategy when producing JAR/JVM files, in particular reordering
the binary order of several definitions (e.g., methods).
For the same case as above, the \bsdiff patches grow from 3.5 KB  to 531~KB,  if provided with the JRE {\sf u10} JAR file
and the  {\sf u11} JAR file generated by \aspa (AST-equivalent to the {\sf u11} JRE version, hence inducing the same \aspa patch).


\psection{Timings}
The experiments for our case-study ran on a 2.5 GHz Intel Core i5 machine with $4$~GB of memory.  We measured the times for \aspa and \bsdiff patch derivation and application. For \aspa we measured an average time of~$60.4$ seconds (s) for
patch derivation and~$7.2$~s for patch application.
As for \bsdiff, the average times were~$31.0$~s for patch derivation and~$1.2$s for patch application.

The times for patch application are specially 
relevant, as they will determine the benefit of transmitting compact
patches over a network vs.\ the option of transmitting full software
versions. Given that the compressed size of \rtjar can be about~$13$ MB using {\sf bzip2} compression
(the JAR files in a JRE distribution contain class files in uncompressed JVM format, since compression is only
applied over the entire JRE distribution bundle), a $7.2$~s download time (the average time to apply a \aspa patch) would be feasible with a $1.8$~MB/s ($14.4$~Mbits/s) download rate. 

The observation above signals a concern for subsequent improvement in {\sf aspa}. The  time for patch application
reflects the prototype stage of the tool, particularly in regard to  I/O implementation details. A great proportion of the time (approx.\ 85\%, 6.1 out of~7.2~s) is consumed by \aspa on I/O operations producing the target JAR file,  
while deriving the AST representation from the source JAR file and changing it through a patch take considerably less longer (approx.\ the other $15$\%).

\vspace{-0.3cm}
\section{Discussion}\label{sec:conclusion}
\vspace{-0.2cm}

\psection{Summary}
We have proposed a methodology for deriving patches for Java software upgrades, based on an AST-level representation of JVM class files,
implemented by the \aspa software tool.
The J2SE API case-study demonstrates the  effectiveness and flexibility of the approach: \aspa patches were found to be significantly smaller 
than patches generated by state-of-the-art binary patching tools, 
and are insensitive to binary-level changes that do not 
correlate with actual changes in Java source code.

\psection{Other languages and compilation formats}
Our proposal may naturally generalize to other languages and compilation formats. For instance, 
other virtual-machine based languages like C\# or Python 
are in principle quite amenable to our methodology.
We can also think of Java again, but considering  the Dalvik bytecode format~\cite{dalvik} that is used in Android devices.
This is a an interesting direction for future work, as Android apps
are updated frequently and in full, leading
to long upgrade times and bandwidth consumption charges for the end user. A more complex scenario is  that 
of programs compiled onto native code, e.g., C/C++ programs. 
In this case it may be harder, in principled
or technical terms, to derive patches  
that relate to source code changes in fine-grained manner.
Even so,  tools like Courgette~\cite{courgette} demonstrate
that factoring out some irrelevant (even if low-level)
differences in the compilation format can lead to much smaller binary patches.

\psection{AST differential analysis} 
The core trait of our methodology is the use of a high-level AST
representation and an associated differential analysis.
We are not aware of previous work that considers
the derivation of patches for \emph{compiled} programs in AST-driven manner with the specific intent of enabling software upgrades.
The AST-difference approach has been employed however for empirical analysis of software evolution~\cite{jdiff,astdiff,xie},
a type of application for which we aim to extend \aspa in the future,
or to derive patches for data formats like XML~\cite{xml-diff}.
We also consider that AST-based software patches may in particular provide an appropriate abstraction level for analysis in 
special and complex contexts, like 
differential symbolic execution~\cite{diffSE} or 
dynamic software updates~(DSU)~\cite{ksplice,dsuc,javau}.
For instance, in the case of DSU, patches are
applied at runtime and a  fine-grained analysis is required 
over the extent and type of changes to decide if and  how a 
patch can be applied during the execution of a program.

\psection{Modular software evolution}
In a broader sense, the problem approached by this paper
relates to principled and modular change analysis
in a software evolution context, in line
with past work by the author in this vein~\cite{apres,acsd}.
The general underlying concern is to attain principled abstractions
for software evolution, such that we can reason on
a modular relation between changes 
to a software artifact and their impact.

{\footnotesize
\psection{Acknowledgements}
The author wishes to thank the anonymous reviewers for their
comments on the draft version of this paper, and Jo\~{a}o Sousa
for help and encouragement. 
This work has been partially funded by 
the Large-Scale Informatics Systems Laboratory (LaSIGE)
at Faculdade de Ci\^{e}ncias da Universidade de Lisboa. 
}
{\footnotesize 
\bibliographystyle{acm}
\bibliography{hotswup}
}


\end{document}